\documentclass[twocolumn,epjc3]{svjour3}          

\RequirePackage[T1]{fontenc}

\smartqed  

\RequirePackage{graphicx}
\RequirePackage{mathptmx}      
\RequirePackage{flushend}
\RequirePackage[numbers,sort&compress]{natbib}
\RequirePackage[colorlinks,citecolor=blue,urlcolor=blue,linkcolor=blue]{hyperref} 

\usepackage{float}
\usepackage{graphicx}
\usepackage{amsmath}
\usepackage{dblfloatfix}
\usepackage{multicol}
\usepackage{color}
\usepackage{widetext}
\usepackage{amsfonts} 
\usepackage{multirow}
\usepackage{tensor}
\usepackage{hyperref}
\journalname{Eur. Phys. J. C}

\begin{document}

\title{Observationally Constrained Cosmological model in $f(Q,\mathcal{L}_{m})$ Gravity with $H(z)$ parameterization }

\author{Vinod Kumar Bhardwaj \thanksref{addr1,e1}, Maxim Khlopov \thanksref{addr2,e2}, Maxim Krasnov \thanksref{addr3,e3}, Saibal Ray \thanksref{addr4,e4}  }

\thankstext{e1}{e-mail: dr.vinodbhardwaj@gmail.com}
\thankstext{e2}{e-mail: khlopov@apc.in2p3.fr}
\thankstext{e3}{e-mail: morrowindman1@mail.ru}
\thankstext{e4}{e-mail: saibal.ray@gla.ac.in}

\institute{Department of Mathematics, GLA University, Mathura 281406, Uttar Pradesh, India \label{addr1} 
\and Virtual Institute of Astroparticle physics, Paris 75018, France \label{addr2} 
\and National Research Nuclear University “MEPHI”, Moscow 115409, Russia \& Institute of Physics, Southern Federal University, Stachki 194, Rostov on Don 344090, Russia \label{addr3} 
\and Centre for Cosmology, Astrophysics and Space Science (CCASS), GLA University, Mathura 281406, Uttar Pradesh, India \label{addr4}}

\date{Received: date / Accepted: date} 

\maketitle

\begin{abstract}
In the present work, we explore an observationally constrained cosmological model in the framework of $f(Q,\mathcal{L}_{m})$ gravity, where $Q$ denotes the non-metricity scalar and $\mathcal{L}_{m}$ represents the matter Lagrangian density. To derive the modified Friedmann field equations, we consider a flat FLRW space-time. We have considered a specific parameterization of the Hubble parameter $H(z)$ to explore the cosmic evolution, which successfully describes the shift of the cosmos from its initial decelerated expansion period to the current accelerated scenario. The free model parameters are constrained using recent observational datasets including Cosmic Chronometers (CC), Pantheon+SH0ES, Union 3.0, DESI-BAO, and CMB distance priors using MCMC approach through the $\chi^2$-minimization process. The derived results indicate that the present model remains consistent with recent cosmological observations. We note that the deceleration parameter exhibits a signature flipping behavior at transition redshift $z_t \approx 0.643$, confirming the transition from matter-dominated deceleration to dark-energy-driven acceleration. The equation of state (EOS) parameter remains in the quintessence region and exhibits an asymptotical approach to the $\Lambda$CDM limit at late times. Moreover, the estimated cosmic age can be found as $13.724^{+0.087}_{-0.048}$ Gyr, which agrees well with recent observational estimations. The statefinder and Om diagnostics support the quintessence nature of the model. At the same time, the examination of energy conditions reveals that two specific energy conditions, viz. Null Energy Condition (NEC) and Dominant Energy Condition (DEC) are fulfilled, while the Strong Energy Condition (SEC) is violated, validating the accelerated expansion of the universe. Therefore, the proposed $f(Q,\mathcal{L}_{m})$ cosmological model offers a viable framework for describing the accelerated evolution of the late universe.
\end{abstract}


\section{Introduction}\label{1}
The mystery of an expanding universe is the cornerstone of modern cosmology. It suggests that the fabric of space-time itself is stretching, pushing galaxies away from one another over time. Einstein's theory of General Relativity (GR) is the theoretical foundation for an expanding universe \cite{Einstein1917gr}. In 1915, Einstein's field equations described how matter and energy curve space-time \cite{Einstein1915eqn}. Theoretical models remained speculative until 20th-century observations provided definitive proof that the universe is not static. In this sequence, the findings of accelerated expansion of the late universe via observations of Ia Supernovae (SN Ia) \cite{Riess1998sn,Perlmutter1999sn}, Cosmic Microwave Background (CMB) \cite{Bennett2003cmb,Spergel2003cmb,Caldwell2004}, Baryon Acoustic Oscillations (BAO) as well as Dark Energy Spectroscopic Instrument (DESI) \cite{Eisenstein2005bao,Percival2010bao}, and large-scale structure surveys (LSS) \cite{Komatsu2009,Riess2004,Koivisto2006} has become one of the most significant developments in modern cosmology. This accelerated behavior suggests the existence of an exotic component known as dark energy (DE), which contributes negative pressure to the cosmic fluid. Theoretically various models are developed in literature for explaining the influence of dark energy in expansion of cosmos, where $\Lambda$ is taken as a substitute of dark energy in Einstein's equation of GR \cite{Peebles20003,Riess2007lambda,Overduin1998lambda,Sahni2000lambda}. Although the standard $\Lambda$CDM model successfully explains a wide range of observational data, it still suffers from several theoretical issues such as the fine-tuning and cosmic coincidence problems \cite{Copeland2006dark}. These shortcomings motivate the search for alternative theories of gravity capable of explaining the present cosmic acceleration without introducing unknown exotic matter components.

So, several alternative theories are proposed to explain this mystery of late time cosmic accelerated expansion by modifying the Einstein-Hilbert action rather adding an additional term as the substitute of DE. These modified gravity theories have emerged as promising candidates to describe the dynamics of the universe beyond GR. In this sequence, to address the problem of cosmological constant, the accelerated phase of the late-time cosmos and other issues eventually gave birth to gravity theories like $f(R)$, where $R$ is the Ricci scalar within the gravitational action, for the basic level \cite{Buchdahl1970,Kerner1982} and even for much advanced cosmological scenarios \cite{Capozziello2006,Amendola2007,CDTT2004,Felice2010,Starobinsky2007,Sotiriou2010,Nojiri2011}.

Later on a specified gravity theory has been proposed and applied by several scientists \cite{Harko2011,Houndjo2012,Moares2017,Bhardwaj2018,Deb2019a,Deb2019b,Maurya2021,Sahoo2022,Koussour2024,Myrzakulov2025pla} on the assumption that the matter Lagrangian is a function of the Ricci scalar ($R$) and the trace of the energy-momentum tensor (EMT) ($T$). The main ideas behind this $f(R,T)$ is to consider the cosmological as well as astrophysical behavior of the universe: (i) in the context of the Big Rip, which concerns about the future singularities, and (ii) to explore the possibility of particle generation by taking into account the quantum mechanical effect. A few more articles related to constraints on energy conditions and stability criteria of Power Law solutions \cite{Sharif2013energy} as well as a non-singular bouncing cosmology \cite{Singh2023non-singular} are available in the literature under the framework of $f(R,T)$ gravity.

Other than $f(R,T)$, a more workable variety of $f(R)$ is $f(R, L_{m})$ gravity theory \cite{ref35,ref36,ref37,ref38,ref39,ref40,ref41,ref42,R10,R4,R9,Romanshu2025,Bhardwaj2025quad} in Riemann space which becomes effective due to the following attributing factors: (i) involves to  connect the `matter Lagrangian density ($L_{m}$)' with an arbitrary function of `Ricci scalar ($f(R)$)' \cite{ref35}, (ii) the non-minimal interaction could provide an extra force, which acts orthogonal to the four-velocity, causing massive particles to deviate from geodesic motion \cite{ref36}, (iii) such matter-geometry couplings lead to important astrophysical implications \cite{ref37,ref38,ref39,ref40,ref41}, (iv) the theory permits possible violations of the equivalence principle, although these effects are tightly constrained by Solar System observations \cite{ref43,ref44}, (v) it is conversant with the energy conditions \cite{ref45} and incorporating non-minimal matter-geometry interactions \cite{ref46}, and (vi) an appreciable comoving entropy is produced in the process of cosmic evolution \cite{Romanshu2025,ref47,ref48}.

Among them, theories based on curvature, torsion, and non-metricity have attracted considerable attention. Recently, symmetric teleparallel gravity, where gravity is characterized by the non-metricity scalar $(Q)$, has provided a geometrically rich framework for cosmological investigations. In this approach, curvature and torsion vanish identically, and gravitational effects are entirely described by non-metricity. Extensions of this framework, particularly $f(Q)$ gravity and its generalized forms, have shown commendable successes to provide adequate explanation of the late-time acceleration of the universe. In this direction, $f(Q,L_m)$ gravity theory is an important extension of symmetric teleparallel gravity, where the gravitational action depends on both the non-metricity scalar $(Q)$ and the `matter Lagrangian density $(L_m)$' \cite{Myrzakulov2024a,Myrzakulov2024b,Myrzakulov2025pdu}. The coupling between geometry and matter introduces new dynamical effects and leads to a non-conservation of the energy-momentum tensor. This non-minimal matter-geometry coupling opens new possibilities for understanding cosmic evolution and dark energy phenomena. Recent studies have demonstrated that $f(Q,L_m)$ gravity can successfully describe several cosmological scenarios, including accelerated expansion and dark energy dominated phases \cite{Samaddar2025,Hazarika2025}.

Inspired by these developments, we examine a cosmological scenario in $f(Q,L_m)$ theory through a parameterized expression of the Hubble parameter $H(z)$. The adopted parameterization allows a smooth transition from an early decelerating universe to the present accelerating phase. To examine the observational viability of the model, we constrain the free parameters using the latest observational datasets, namely CC, Pantheon+SH0ES, Union 3.0 Supernova compilation, DESI-BAO measurements, and CMB distance priors through the standard $\chi^2$-statistical analysis applying the MCMC analysis. 

Additionally, we study several important cosmological quantities, viz. the energy density, EOS parameter, age, statefinder diagnostic, Om diagnostic, and other important physical parameters. For checking the physical viability of the model, we have also analyzed energy conditions. The results obtained suggest that the derived model effectively explains the present accelerated expansion of the universe and remains consistent with recent observational findings.

The paper is organized as follows. In Section 2, we present the basic formalism and field equations of $f(Q,L_m)$ gravity. Section 3 discusses on the observationally available datasets with statistical analysis as used to constrain the model parameters. In Section 4, we analyze the cosmological behavior of the model through various diagnostics and physical parameters. Section 5 is devoted to the discussion of energy conditions. Finally, Section 6 summarizes the main findings and conclusions of the present investigation.

\section{Metric and field equations in $f(Q,\mathcal{L}_{m})$ theory of gravity}\label{2}

\subsection{Background equations}\label{2.1}
In this segment, we begin with a brief discussion of the geometric principles that form the basis of gravitational theories, emphasizing the role of the general space-time line element. We then present the action corresponding to the ($f(Q,\mathcal{L}_{m})$) theory of gravity and derive the associated gravitational field equations through the application of the variational principle, thereby providing deeper insight into gravitational phenomena within this geometric framework. Moreover, we investigate the non-conservation issue of the matter energy-momentum tensor, which illustrates the effects arising from the coupling in between the underlying geometry and allied the matter Lagrangian.

The geometrical interpretation of gravity using Riemann tensor is given by
\begin{equation}\label{e1}
{R^{\alpha}}_{\beta \mu \nu} = \partial_{\mu} {X^{\alpha}}_{ \nu \beta}-\partial_{\nu} {X^{\alpha}}_{ \mu \beta}+{X^{\alpha}}_{ \mu \lambda} {X^{\lambda}}_{ \nu \beta}-{X^{\alpha}}_{ \nu \lambda} {X^{\lambda}}_{ \mu \beta}.
\end{equation}

The Riemann curvature tensor along with its various contractions are derived through affine connection, which provides the mathematical framework for characterizing the curvature of spacetime. According to Weyl-Cartan geometry which is generalization of Riemann geometry, the affine connection ${X^{\alpha}}_{\mu \nu}$ consists of three components: Levi-Civita symmetric connection ${\varGamma^{\alpha}}_{\mu \nu}$, the contortion tensor ${K^{\alpha}}_{\mu \nu}$ describing unsymmetrical contribution, and the dis-formation tensor ${L^{\alpha}}_{\mu \nu}$, which represents the effects arising from non-metricity
\begin{equation}\label{e2}
{X^{\alpha}}_{\mu \nu} = {\varGamma^{\alpha}}_{\mu \nu}+{K^{\alpha}}_{\mu \nu}+{L^{\alpha}}_{\mu \nu}.
\end{equation}

The torsion-free Levi-Civita connection ${\varGamma^{\alpha}}_{\mu \nu}$ describes the curvature and parallel transport in a metric-compatible space. It is equivalent to the second order Christoffel symbol encapsulating the gravitational effect in GR and is defined according to   
\begin{equation}\label{e3}
{\varGamma^{\alpha}}_{\mu \nu} = \frac{1}{2} g^{\alpha \lambda} \left(\partial_{\mu} g_{\lambda \nu}  +\partial_{\nu} g_{\lambda \mu}-\partial_{\lambda} g_{\mu \nu}\right). 
\end{equation}

The contortion tensor ${K^{\alpha}}_{\mu \nu}$ in terms of torsion tensor ${T^{\alpha}}_{\mu \nu}$ is expressed as
\begin{equation}\label{e4}
{K^{\alpha}}_{\mu \nu} = \frac{1}{2} \left( T\indices{_{\mu}^{\alpha}_{\nu}}+T\indices{_{\nu}^{\alpha}_{\mu}}+T\indices{^{\alpha}_{\mu \nu}} \right). 
\end{equation}

The lack of symmetry in an affine connection can be quantified by the torsion tensor which essentially implies the parallel transport of a vector around a closed loop.

In contrast, the dis-formation tensor $L^{\alpha}_{\mu \nu}$ is associated with the expansion or contraction properties of space-time. During parallel transport, the magnitude of a vector can vary along its trajectory, and this change in length is characterized by the non-metricity tensor
\begin{equation}\label{e5}
L^{\alpha}_{\mu \nu} = \frac{1}{2} \left(Q\indices{^{\alpha}_{\mu \nu}} - Q\indices{_{\mu}^{\alpha}_{\nu}} - Q\indices{_{\nu}^{\alpha}_{\mu}}\right). 
\end{equation}

Now $Q\indices{_{\alpha}_{\mu}_{\nu}}$, the non-metricity tensor, can be defined as \cite{Jimenez2018}
\begin{equation}\label{e6}
Q\indices{_{\alpha}_{\mu}_{\nu}}= \nabla \indices{_\alpha} g\indices{_{\mu}_{\nu}} = \partial_{\alpha} g\indices{_{\mu}_{\nu}}- {X \indices{^\beta}} \indices{_\alpha _\mu} g\indices{_\beta _\nu} - {X \indices{^\beta}} \indices{_\alpha _\nu} g\indices{_\mu _\beta}.
\end{equation}

In the theories of metric-affine gravity, the non-metricity conjugate superpotential $P \indices{^{\alpha}_\mu _\nu}$ is defined to introduce the boundary term that appears in the action \cite{Xu2019} and is given by
\begin{equation}\label{e7}
P\indices{^{\alpha}_\mu _\nu} = -\frac{1}{2} L \indices{^{\alpha}_\mu _\nu} +\frac{1}{4} \left(Q^{\alpha} -\tilde{Q}^{\alpha}\right) g\indices{_{\mu}_{\nu}}-\frac{1}{4} \delta^{\alpha} \left(\indices{_\mu} Q\indices{_\nu} \right),
\end{equation}
where $Q^\alpha = Q\indices{^{\alpha}_{\mu} ^{\mu}}$ and $\tilde{Q}^\alpha = Q\indices{_{\mu}^{\alpha \mu}}$ are the non-metricity vectors. By contracting the super-potential tensor with the non-metricity tensor, the non-metricity scalar can be obtained as
\begin{equation}\label{e8}
Q = - Q\indices{_{\lambda}_\mu _\nu}  P\indices{^{\lambda}^\mu ^\nu}.
\end{equation}

For the $f(Q,\mathcal{L}_{m})$ gravity, the action principal to derive a set of field equations is defined as 
\begin{equation}\label{e9}
S = \int f(Q,\mathcal{L}_{m}) \sqrt{-g} \, d^{4}x, 
\end{equation}
where $\sqrt{-g}$ corresponds to the determinant of the metric tensor, and $f(Q,\mathcal{L}_{m})$ signifies a general function involving both the non-metricity scalar ($Q$) and the density $\mathcal{L}_{m}$ of the matter Lagrangian.\\

By taking the variation of the action (\ref{e9}) with the metric tensor $g_{\mu \nu}$, the following field equation in $f(Q,\mathcal{L}_{m})$ gravity can be derived as
\begin{eqnarray}\label{e10}
&\frac{2}{\sqrt{-g}} \nabla_{\alpha} \left(f_{Q} \sqrt{-g} P\indices{^{\alpha}_{\mu}_{\nu}}\right)+f_{Q} \left( P\indices{_{\mu}_{\alpha}_{\beta}} Q\indices{_{\nu}^{\alpha}^{\beta}}-2  Q^{\alpha \beta}_{\mu} \, P\indices{_{\alpha}_{\beta}_{\nu}}\right)+\frac{1}{2} f g_{\mu \nu}\nonumber\\
&= \frac{1}{2} f_{\mathcal{L}_{m}} \left(g_{\mu \nu} \mathcal{L}_{m}-T_{\mu \nu}\right),
\end{eqnarray}
where $f_Q\equiv\frac{\partial f}{\partial Q}$, $f_{\mathcal{L}_{m}}\equiv\frac{\partial f}{\partial \mathcal{L}_{m}}$, and $T_{\mu \nu}$ represents the energy-momentum tensor for matter, defined as 
\begin{small}
\begin{equation} \label{e11}
T_{\mu \nu}=-\frac{2}{\sqrt{-g}} \frac{\delta (\sqrt{-g} \mathcal{L}_{m})}{\delta g^{\mu \nu}} = g^{\mu \nu} \mathcal{L}_{m} - 2 \frac{\delta \mathcal{L}_{m}}{\delta g^{\mu \nu}}.
\end{equation}
\end{small}

By varying the gravitational action, with respect to the connection, one can get the field equations as
\begin{small}
\begin{equation}\label{e12}
\nabla_{\mu} \nabla_{\nu} = \left(4 \sqrt{-g} f_{Q} P \indices{^{\mu \nu}_{\alpha}}+H \indices{_{\alpha}^{\mu \nu}}\right)=0, 
\end{equation}
\end{small}
where $H \indices{_{\alpha}^{\mu \nu}}$ represents the hyper-momentum density, defined by $H \indices{_{\alpha}^{\mu \nu}}=\sqrt{-g} f_{\mathcal{L}_{m}} \frac{\delta \mathcal{L}_{m}}{\delta X\indices{^{\alpha}_\mu_\nu}}$.

Therefore, applying the covariant derivative to the field equation (\ref{e10}), one can derive 
\begin{small}
\begin{eqnarray} \label{e13}
&D_{\mu} T\indices{^{\mu}_{\nu}} = \frac{1}{f_{\mathcal{L}_{m}}}\bigg[\frac{2}{\sqrt{-g}} \nabla_{\alpha} \nabla_{\mu} H\indices{_{\nu}^{\alpha\mu}}+\nabla_{\mu} A\indices{^{\mu}_{\nu}}-\nabla_{\mu}\left(\frac{1}{\sqrt{-g}} \nabla_{\alpha} H\indices{_{\nu}^{\alpha\mu}}\right)\bigg]\nonumber\\
&=B_{\nu}\neq 0. 
\end{eqnarray}
\end{small}

From Eq. (\ref{e13}), it is evident that the matter EMT is not conserved under $f(Q, \mathcal{L}_{m})$ gravity framework. The corresponding non-conservation tensor $B_\nu$ depends on several dynamical quantities, including the non-metricity scalar $Q$, the matter Lagrangian density $\mathcal{L}_{m}$, and various thermodynamic constraints.

If we consider matter as a perfect fluid with energy density $\rho$ and pressure $p$, the EMT can be defined as 
\begin{equation}\label{e14}
T\indices{_{\mu}_{\nu}} = (\rho+p) u_{\mu} u_{\nu} + p g_{\mu \nu},
\end{equation}
where $u_\mu$ is the fluid-four velocity vector, with components $u^\mu = (1, 0,0,0)$.

To study the cosmological evolution in $f(Q, \mathcal{L}_{m})$ gravity, we consider an isotropic and homogeneous universe described by  
flat Friedmann-Lema\^{i}tre-Robertson-Walker (FLRW) space-time metric:
\begin{equation} \label{e15}
ds^2 = a^{2}(t) \left(dx^2+dy^2+dz^2\right)-dt^2,
\end{equation}
where $ a(t)$ is the scale factor. The expansion rate of the universe is defined by the Hubble parameter $H = \frac{\dot{a}}{a}$. For a universe filled with perfect fluid, the Lagrangian density for cosmic matter can be expressed as either $\mathcal{L}_{m} = p$ or $\mathcal{L}_{m} = -\rho$. Hence, in the comoving coordinate system, the non-zero components of the energy-momentum tensor can be provided as $T_{\mu \nu} = (-\rho, p, p, p)$.

For perfect matter fluid, the modified Friedmann equations describing the dynamics of the FLRW universe in $f(Q, \mathcal{L}_{m})$ gravity are derived as \cite{Hazarika2024,Myrzakulov2025modified}
\begin{equation} \label{e16}
3 H^2 = \frac{1}{4 f_Q} \bigg[f-f_{ \mathcal{L}_{m}} (\rho+ \mathcal{L}_{m})\bigg], 
\end{equation}
\begin{equation} \label{e17}
\dot{H}+3 H^2 = -\frac{\dot{f_Q}}{f_Q}+\frac{1}{4 f_Q} \bigg[f+f_{ \mathcal{L}_{m}} (p- \mathcal{L}_{m})\bigg]. 
\end{equation}

The continuity equation within the framework of $f(Q, \mathcal{L}_{m})$ gravity can be obtained as \cite{Harko2018Qgrav}
\begin{equation} \label{e18}
\dot{\rho}+3 H (\rho+p) = B_{\mu} u^{\mu}
\end{equation}
where $B_{\mu} u^{\mu}$ is the term associated with dissipation or generation of energy. The system follows the conservation laws associated with standard gravitational theories when $B_{\mu} u^{\mu}=0$, and energy processes become dominant if $B_{\mu} u^{\mu} \neq 0$.

Accordingly, the modified Friedmann equations can be written via an effective energy density and pressure as follows:
\begin{equation} \label{e19}
3 H^2 = \rho_{eff},  \, \, \, \,  2\dot{H}+3 H^2 = -p_{eff},
\end{equation}
where effective energy density and pressure are defined as
\begin{equation} \label{e20}
\rho_{eff} = \frac{1}{4 f_Q} \bigg[f-f_{ \mathcal{L}_{m}} (\rho+ \mathcal{L}_{m})\bigg], 
\end{equation} 
\begin{equation} \label{e21}
p_{eff} = 2\frac{\dot{f_Q}}{f_Q} H-\frac{1}{4 f_Q} \bigg[f+f_{ \mathcal{L}_{m}} (\rho+2p- \mathcal{L}_{m})\bigg]. 
\end{equation}

From Eq. (\ref{e19}), the generalized effective conservation equation in theory of $f(Q, \mathcal{L}_{m})$ gravity can be recasts as
\begin{equation} \label{e22}
\dot{\rho_{eff}}+3 H (\rho_{eff}+p_{eff}) = 0. 
\end{equation}

From Eq. (\ref{e19}), the deceleration parameter can be expressed as
\begin{equation} \label{e23}
q = \frac{1}{2}+\frac{3}{2} \frac{p_{eff}}{\rho_{eff}}. 
\end{equation}

\subsection{Cosmological Model}\label{2.2}
In the present study, we consider a generalized functional form for the $f(Q, \mathcal{L}_{m})$ gravity framework which can be given as \cite{Hazarika2024}
\begin{equation} \label{e24}
f(Q, \mathcal{L}_{m}) = -\alpha Q + 2 \mathcal{L}_{m} +\beta,
\end{equation}
with $\mathcal{L}_{m} = \rho$, where $\alpha$ and $\beta$ are constants and hence the Friedmann equations reduces to
\begin{equation} \label{e25}
3 H^2 = \frac{\rho}{2 \alpha} -\frac{\beta}{2 \alpha}, 
\end{equation}

\begin{equation} \label{e26}
2\dot{H}+3 H^2 = - \frac{p}{\alpha} -\frac{\beta}{2\alpha}. 
\end{equation}

Specifically, for $\alpha=1$ and $\beta=0$, the standard Friedmann equations of GR can be recovered. We consider the specific parameterization of the Hubble parameter in the following form of redshift function to explore the dynamics universe in framework of $f(Q, \mathcal{L}_{m})$ gravity:
\begin{equation} \label{e27}
H(z) = H_0 \left(\frac{\delta +\gamma  (z+1)^{\eta }}{\gamma +\delta }\right)^{\frac{3}{2 \eta }},
\end{equation}
where $H_0$ is the present value of Hubble parameter.

\section{Observational Data and Statistical Analysis}\label{3}
In the present segment, we constrain the free parameters of the derived cosmological model by using the minimum $\chi^2$ process based on the observational data sets. The best-fit values of the parameters $\delta$, $\gamma$, $\eta$, and $H_{0}$ are extracted by analyzing the available data via the successive operational procedures which are as follows: :

\textbf{1. CC}: This data sample consists of 31 observational measurements of the Hubble parameter $H(z)$ compiled by the cosmic chronometric technique within the redshift range $[0.07, 1.965]$ \cite{Moresco2015,Sharov2018,Yu2018}.

\textbf{2. Pantheon+SH0ES}: We employ the latest Pantheon compilation of SN Ia observations spanning the redshift interval $[0.001, 2.26]$ \citep{conley2010supernova, Scolnic2018}. This dataset combines SN Ia measurements from 18 independent surveys, with all light curves consistently reanalyzed using the SALT2 fitting framework \citep{Brout2022}.

\textbf{3. BAO-DESI data}: We utilize the BAO-DESI datasets, which comprises 12 BAO measurements obtained from findings of different tracers by DESI  \citep{adame2025desi,lodha2025extended,abdul2025desi}. In the measurements we have included the following datasets $d_\mathrm{M}(z)/r_\mathrm{d}$, $d_\mathrm{H}(z)/r_\mathrm{d}$, and $d_\mathrm{V}(z)/r_\mathrm{d}$, where $d_\mathrm{M}(z)= d_\mathrm{L}(z)/(1+z)$ denotes the transverse comoving distance, $d_\mathrm{H}(z)= c/H(z)$ represents the Hubble distance, and $d_\mathrm{V}(z)=\left[z d_\mathrm{H} (z) d^2_\mathrm{M}(z)\right]^{1/3}$ corresponds to the spherically averaged distance measure. Here, $r_\mathrm{d}$ is the sound horizon at the baryon drag era serving as a cosmological standard rule maker.

\textbf{4. Union 3.0}: Here we utilize the latest Union 3.0 compilation, which is created from the recent SN Ia observations. This dataset represents the largest homogeneous SN Ia sample to the present epoch which combines 2087 supernovae within the redshift interval $0.01 < z < 2.26$ from 24 independent surveys through a unified Bayesian framework \citep{rubin2025union}. It is important to note that our analysis employs the publicly available binned version of this dataset, which is expressed in terms of distance moduli. Although the original observations are reported in terms of the apparent magnitude ($m_B$), the corresponding distance modulus is defined as:
\begin{equation} \label{e28}
\mu (z) \equiv m_\mathrm{B} (z) - M_\mathrm{B} = 5 \bigg[5 + \log_{10} \bigg(\frac{d_\mathrm{L} (z)}{\mathrm{Mpc}}\bigg)\bigg],
\end{equation}
where the luminosity distance $d_\mathrm{L}$ is defined as
\begin{equation} \label{e29}
d_\mathrm{L} (z) = c (1+z) \int_{0}^{z} \frac{dz'}{H(z')}.
\end{equation}

Although there are some studies that suggest the luminosity, equivalently the absolute magnitude, of Type Ia supernovae remains independent of cosmic redshift \citep{tutusaus2019model}. Consequently, the SN Ia dataset is considered to be characterized by an average absolute magnitude of $M_B = -19.22$ \citep{benevento2020can}.

\textbf{5. CMB}: We utilize the CMB distance priors reported by \cite{chen2019distance}, which were extracted from the Planck 2018 temperature and polarization observations and also are included the TT, TE, EE, and lowE datasets \citep{aghanim2020planck}. These priors effectively capture key geometrical information contained in the full CMB power spectrum through parameters such as the acoustic scale ($l_\mathrm{A}$), the shift parameter ($R_\mathrm{shift}$) and the baryon density ($\Omega_\mathrm{b}h^2$). The use of these quantities enables the incorporation of CMB limitations into the analysis without the need to perform a complete treatment of cosmological perturbations.

Accordingly, the CMB distance prior dataset can be characterized by the aforementioned parameters, namely $R_\mathrm{shift}$ and $l_\mathrm{A}$, whose expressions are given below: 
\begin{equation} \label{e30}
R_\mathrm{shift} = \frac{(1+z_{*}) D_\mathrm{A} (z_{*})\sqrt{\Omega_m H^2_{0}}}{c}, ~l_\mathrm{A} =(1+z_{*}) \frac{\pi D_\mathrm{A} (z_{*})}{r_\mathrm{s}(z_{*})},
\end{equation}
where $z_{*}$ represents the redshift at the epoch of photons decoupling, while $r_\mathrm{s}(z_{*})$ corresponds to the comoving sound horizon calculated at the same epoch.

The free parameters $\delta$, $\gamma$, $\eta$, and $H_{0}$ of the derived model are constrained by employing observational datasets together with statistical techniques. Here, $E_{\mathrm{obs}}$ denotes the observed values obtained from the data, whereas $E_{\mathrm{th}}$ corresponds to the theoretical predictions of the cosmological model. By statistically comparing $E_{\mathrm{obs}}$ with $E_{\mathrm{th}}$, the best-fit values of the model parameters can be estimated. To achieve the highest level of accuracy in parameter determination, the $\chi^{2}$ minimization method is adopted. If $\sigma$ denotes the standard deviation associated with the observational measurements, the $\chi^{2}$ estimator is defined as follows:
\begin{equation} \label{chi1}
\chi^{2} = \sum_{i=1}^{N} \frac{\left(E_{obs}(z_{i})-E_{th}(z_{i})\right)^2}{\sigma_{i}^{2}},
\end{equation}
where $N$ represents the sample points in the data.

The joint estimator $\chi^{2}_{combined}$ for the model is defined as
\begin{equation} \label{chi2}
\chi^{2}_{combined} = \chi^{2}_{CC} + \chi^{2}_{CMB} +\chi^{2}_{Pantheon+SH0ES} + \chi^{2}_{Union3.0} + \chi^{2}_{DESI}.
\end{equation}

\begin{figure}[ht]
	\centering
	\includegraphics[width=0.5\textwidth]{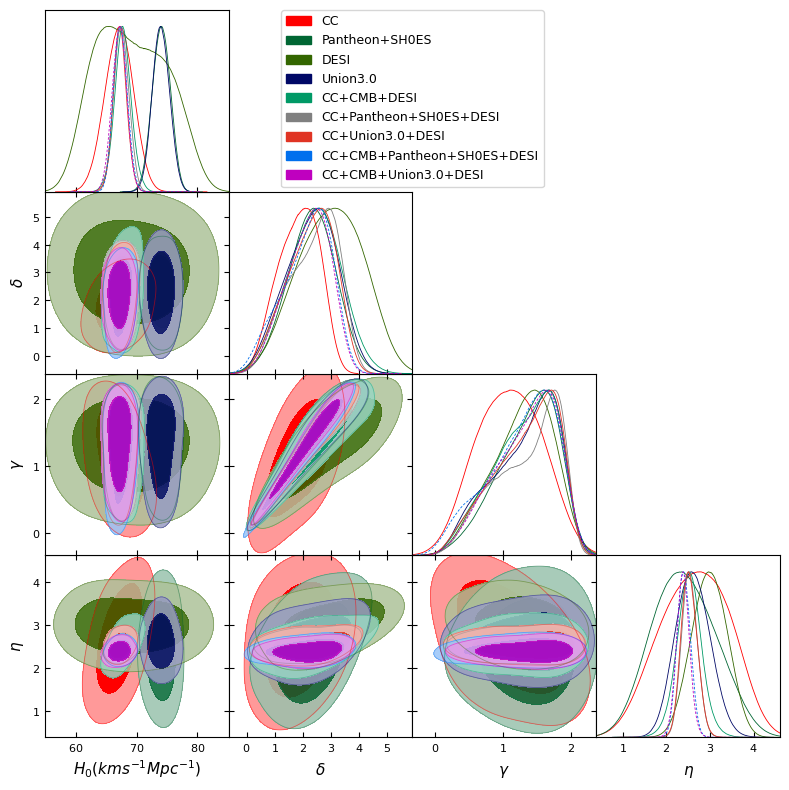}
	\caption{Contour plots for the proposed model with 68$\%$ and $95 \%$ level of confidence.}\label{fig1}
\end{figure}

\begin{table}[h]
	\caption{The estimated parametric values of the specified datasets, where I, II, III, IV, V, VI, VII, VIII, and IX respectively stand for CC, DESI, Pantheon+SH0ES, Union3.0, CC+CMB+DESI, CC+Pantheon+SH0ES+DESI, CC+Union3.0+DESI, CC+CMB+Pantheon+SH0ES+DESI, and CC+CMB+Union3.0+DESI.}
	\begin{center}
		\begin{tabular}{ccccc}
			\hline\hline 
			\tiny Parameters & \tiny $H_{0}$ $(km s^{-1} Mpc^{-1})$ & \tiny $\delta$ & \tiny $\gamma$ & \tiny $\eta$ \\
			\hline
			\tiny I & \tiny $67.114 \pm 2.233$ & \tiny $1.862 \pm 0.834$ & \tiny $1.094 \pm 0.517$ & \tiny $2.662 \pm 0.822$  \\ 
			\hline
			\tiny II & \tiny $69.130 \pm 6.654$ & \tiny $ 2.990 \pm	1.211$ & \tiny $1.297 \pm 0.496$ & \tiny $ 2.990 \pm 0.389$  \\
			\hline
			\tiny III & \tiny $74.074 \pm 1.461$ & \tiny $2.318 \pm 	0.855$ & \tiny $1.370 \pm 0.472$ & \tiny $2.406 \pm 0.748$ \\
			\hline
			\tiny IV & \tiny $74.018 \pm 1.347$ & \tiny $2.312 \pm 	1.024$ & \tiny $1.332 \pm 0.568$ & \tiny $2.606 \pm 0.385$ \\
			\hline
			\tiny V & \tiny $67.737	\pm 1.297$ & \tiny $2.397 \pm	1.028$ & \tiny $1.324 \pm 0.516$ & \tiny $2.523 \pm 	0.261$ \\
			\hline
			\tiny VI & \tiny $67.479 \pm 1.142$ & \tiny $2.393 \pm	1.008$ & \tiny $1.363 \pm 0.569$ & \tiny $2.533 \pm	0.170$ \\
			\hline
			\tiny VII & \tiny $67.244 \pm 1.028$ & \tiny $2.288 \pm	0.903$ & \tiny $1.339 \pm 0.540$ & \tiny $2.526 \pm 0.188$ \\
			\hline
			\tiny VIII & \tiny $67.326 \pm 1.093$ & \tiny $2.152 \pm	0.937$ & \tiny $1.284 \pm 0.588$ & \tiny $2.403 \pm 0.142$ \\
			\hline
			\tiny IX & \tiny $67.134 \pm 1.124$ & \tiny $2.200 \pm 0.881$ & \tiny $1.346 \pm 0.534$ & \tiny $2.383	
			\pm 0.146$ \\
			\hline
			\end{tabular}
	\end{center}\label{tab1}
\end{table}


\section{Analysis of the model's cosmological characteristics} \label{4}
\begin{equation} \label{e33}
\rho = \frac{\beta }{2}+3 \alpha  H^{2}_0 \left(\frac{\delta +\gamma  (z+1)^{\eta }}{\gamma +\delta }\right)^{3/\eta },
\end{equation}

\begin{equation}\label{e34}
p = -\frac{\beta }{2}-\frac{3 \alpha  \delta  H^{2}_0 \left(\frac{\delta +\gamma  (z+1)^{\eta }}{\gamma +\delta }\right)^{\frac{3}{\eta }-1}}{\gamma +\delta },
\end{equation}

\begin{equation} \label{e35}
\omega = \frac{2 \left(-\frac{\beta }{2}-\frac{3 \alpha  \delta  H^{2}_0 \left(\frac{\delta +\gamma  (z+1)^{\eta }}{\gamma +\delta }\right)^{\frac{3}{\eta }-1}}{\gamma +\delta }\right)}{\beta +6 \alpha  H^{2}_0 \left(\frac{\delta +\gamma  (z+1)^{\eta }}{\gamma +\delta }\right)^{3/\eta }}.
\end{equation}

\begin{figure}[ht]
	\centering
	(a) \includegraphics[width=0.5\textwidth]{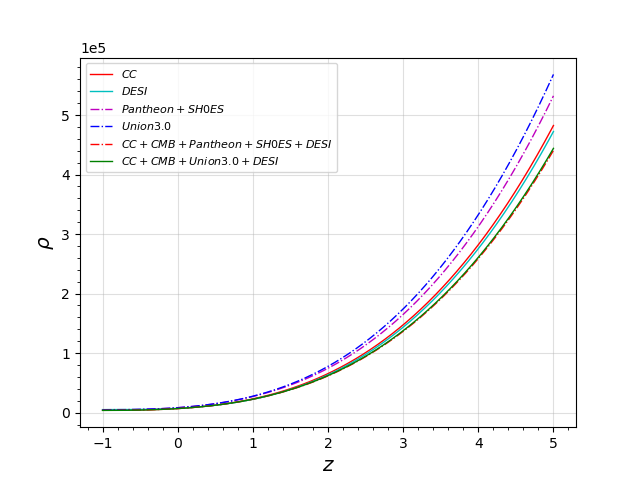}\\
	(b) \includegraphics[width=0.5\textwidth]{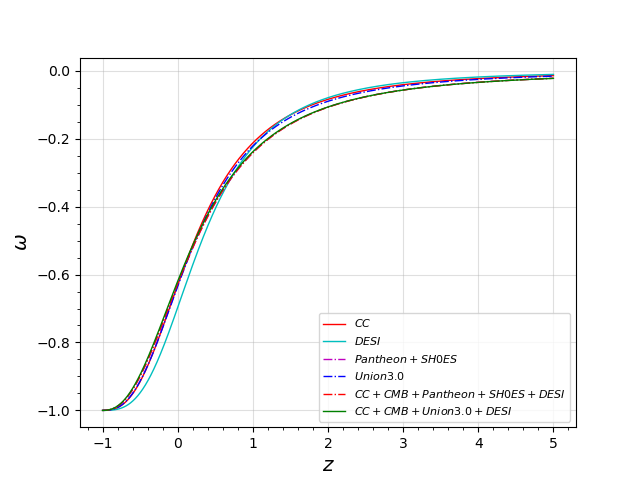}
	\caption{Graphical plots for (a) energy density, (b) trajectory of EOS parameter}. \label{fig2}
\end{figure}

For the presented model within $f(Q,\mathcal{L}_{m})$  gravity theory, the energy density remains positive throughout the cosmic evolution, as illustrated in Fig. \ref{fig2}(a). The equation of state (EOS) parameter characterizes different phases of the universe depending on the value of $\omega$. For $\omega > 0$, it corresponds to the early Big Bang epoch, while $\omega = 0$ represents a matter-dominated universe. Under this situation, dark matter behaves as a non-relativistic component with $\omega = 0$, whereas relativistic matter, such as radiation, is characterized by $\omega = 1/3$. The interval $-1 < \omega \leq 0$ describes the quintessence regime. When $\omega = -1$, the EOS parameter converges to the $\Lambda$CDM scenario, whereas $\omega < -1$ indicates a phantom-like phase. The evolution of the EOS parameter for the derived model is presented in Fig. \ref{fig2}(b), which reveals that the universe initially evolves within the quintessence regime and gradually approaches the $\Lambda$CDM phase at late times.

\subsection{Age of the universe} \label{4.1}
The age of the universe can be calculated by employing the connection between the redshift ($z$)  and the scale factor ($a$), expressed as:
\begin{equation} \label{e36}
dt=-\frac{dz}{(1+z) H(z)}\implies \int_{t}^{t_{0}} dt=-\int_{z}^{0} \frac{1}{(1+z) H(z)} dz. 
\end{equation}

From Eqs. (\ref{e27}) \& (\ref{e36}), the present age of universe ($t_0$) in the derived model is obtained as:    
\begin{small}
\begin{equation}\label{e37}
t_{0}=\lim_{x \to \infty}\int_0^x \frac{1}{ (z+1) H_0 \left(\frac{\delta +\gamma  (z+1)^{\eta }}{\gamma +\delta }\right)^{\frac{3}{2 \eta }}} \, dz.
\end{equation}
\end{small}

\begin{figure}[ht]
	\centering
	(a)\includegraphics[width=0.9\linewidth]{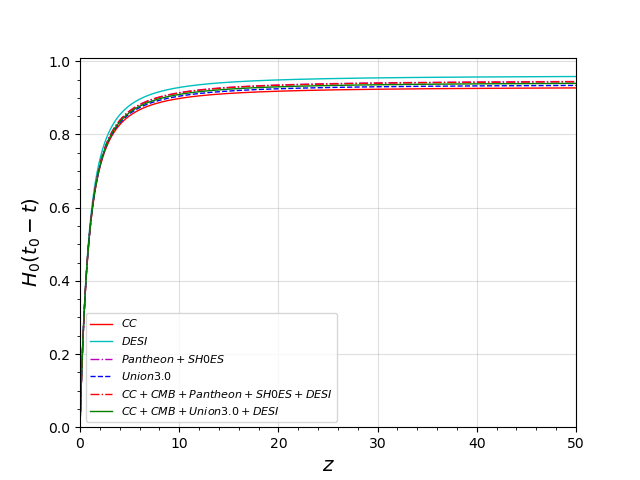}\\
	(b) \includegraphics[width=0.9\linewidth]{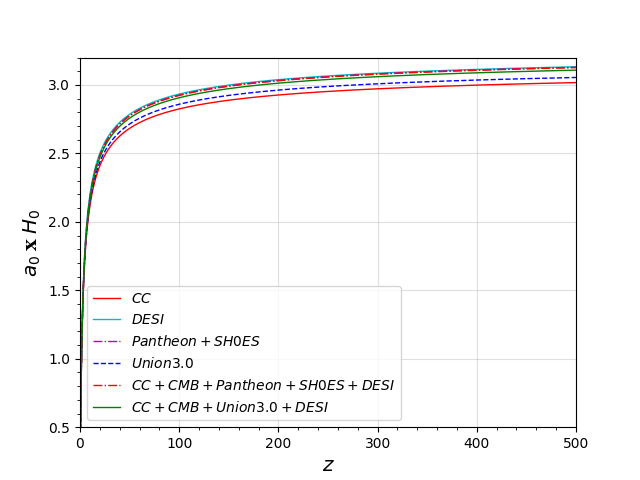}
	\caption{Graphical plots for (a) $ H_{0} (t_{0}-t) $ versus $ z $, (b) $a_{0} x H_{0}$ vs $z$.}\label{fig3} 
\end{figure}

The age of the Universe can be estimated from the above-mentioned relation by plotting $H_{0}(t_{0}-t)$ as a function of the redshift $z$. Figure \ref{fig3} (a) illustrates the behavior of $H_{0}(t_{0}-t)$ for various observational datasets, including CC, CMB, Pantheon+SH0ES, Union3.0, DESI, and their combined datasets. From the analysis of the joint dataset comprising \\
{\small CC+CMB+Union3.0+DESI}, we obtained $H_{0}t_{0}=0.9403$, which corresponds to a present age of the Universe of $13.724^{+0.087}_{-0.048}$ Gyrs. The estimated ages derived from the proposed model for the individual and combined datasets are listed in Table \ref{tab2}. These results are consistent with recent observational estimates reported in \cite{Renzini1996,Masi2002,Hinshaw2013,Bond2013,Bhardwaj2024lyra,Bhardwaj2024scalar}.

\begin{table*}[t]
	\caption{A thorough summary of the numerical results estimated from the proposed $f(Q,\mathcal{L}_{m})$ gravity theory, where I, II, III, IV, V, VI, VII, VIII, and IX stand for CC, DESI, Pantheon+SH0ES, Union3.0, CC+CMB+DESI, CC+Pantheon+SH0ES+DESI, CC+Union3.0+DESI, CC+CMB+Pantheon+SH0ES+DESI, and CC+CMB+Union3.0+DESI, respectively.}
    \begin{center}
		\begin{tabular}{ccccccc}
			\hline\hline 
			\tiny Parameters & \tiny $q_{0}$ & \tiny $z_{t}$ & \tiny $j_{0}$ & \tiny $\omega_{0}$ & \tiny $Age(Gyrs)$ & \tiny Particle Horizon $R_p$ \\
			\hline
			\tiny I & \tiny \tiny $-0.445^{+0.006}_{-0.016}$ & \tiny $0.586^{+0.405}_{-0.171}$ & \tiny  $0.679^{+0.237}_{-0.144}$ & \tiny  $0.630^{+0.004}_{-0.009}$ & \tiny $13.524^{+0.775}_{-0.287}$  & \tiny $3.001^{+0.567}_{-0.273}$.$H^{-1}_{0}$\\ 
			\hline
			\tiny II & \tiny $-0.545^{+0.006}_{-0.011}$ & \tiny $ 0.666^{+0.110}_{-0.086}$ & \tiny $0.790^{+0.094}_{-0.084}$ & \tiny $-0.697^{+0.007}_{-0.004}$ & \tiny $13.584^{+1.099}_{-0.873}$ & \tiny $3.370^{+0.011}_{-0.382}$.$H^{-1}_{0}$\\
			\hline
			\tiny III & \tiny$-0.443^{+0.006}_{-0.015}$ & \tiny $0.659^{+0.373}_{-0.178}$ & \tiny $0.621^{+0.190}_{-0.135}$ & \tiny $-0.629^{+0.010}_{-0.004}$ & \tiny $12.504^{+0.638}_{-0.377}$  & \tiny $3.113^{+0.508}_{-0.277}$.$H^{-1}_{0}$ \\
			\hline
			\tiny IV & \tiny $-0.452^{+0.004}_{-0.008}$ & \tiny $0.611^{+0.123}_{-0.089}$ & \tiny $0.670^{+0.097}_{-0.085}$ & \tiny $-0.635^{+0.005}_{-0.003}$ & \tiny $12.381^{+0.128}_{-0.102}$ & \tiny $3.043^{+0.185}_{-0.142}$.$H^{-1}_{0}$\\
			\hline
			\tiny V & \tiny $-0.468^{+0.008}_{-0.025}$ & \tiny $0.669^{+0.046}_{-0.065}$ & \tiny $0.658^{+0.062}_{-0.067}$ & \tiny $-0.645^{+0.017}_{-0.005}$ & \tiny $13.743^{+0.041}_{-0.156}$  & \tiny $3.119^{+0.080}_{-0.092}$.$H^{-1}_{0}$\\
			\hline
			\tiny VI & \tiny $-0.456\pm 0.001$ & \tiny $0.643^{+0.061}_{-0.049}$ & \tiny $0.654^{+0.039}_{-0.035}$ & \tiny $-0.637^{+0.001}_{-0.001}$ & \tiny $13.702^{+0.059}_{-0.031}$ & \tiny $3.088^{+0.088}_{-0.076}$.$H^{-1}_{0}$ \\
			\hline
			\tiny VII & \tiny $-0.446^{+0.003}_{-0.002}$ & \tiny $0.625^{+0.072}_{-0.056}$ & \tiny $0.649^{+0.041}_{-0.038}$ & \tiny $-0.631^{+0.002}_{-0.002}$ & \tiny $13.662^{+0.054}_{-0.009}$ & \tiny $3.064^{+0.110}_{-0.086}$.$H^{-1}_{0}$\\
			\hline
			\tiny VIII & \tiny $-0.446^{+0.003}_{-0.002}$ & \tiny $0.657^{+0.074}_{-0.056}$ & \tiny $0.618^{+0.028}_{-0.022}$ & \tiny $-0.627^{+0.004}_{-0.007}$ & \tiny $13.743^{+0.056}_{-0.012}$ & \tiny $3.108^{+0.116}_{-0.079}$.$H^{-1}_{0}$\\
			\hline
			\tiny IX & \tiny $-0.430^{+0.002}_{-0.001}$ & \tiny $0.643^{+0.052}_{-0.044}$ & \tiny $0.607^{+0.034}_{-0.028}$ & \tiny $-0.620^{+0.001}_{-0.001}$ & \tiny $13.724^{+0.087}_{-0.048}$& \tiny $3.094^{+0.076}_{-0.066}$.$H^{-1}_{0}$\\
			\hline
		\end{tabular}
	\end{center}\label{tab2}
\end{table*}

\subsection{Particle horizon} \label{4.2}

In cosmology, the particle horizon is an important concept that links the expansion history of the universe to the limits of the observable cosmos. It corresponds to the maximum distance from which light emitted since the beginning of the universe could have reached an observer by the present epoch \cite{Bentabol2013}. Consequently, it represents the most distant region from which information about the early universe can be obtained. Since the expansion rate of the universe has evolved over time, the particle horizon is most appropriately described using proper distance. Let us consider a light signal emitted at cosmic time $t_p$ from a distant source of an expanding universe and propagating along the $x$-direction. If the scale factor at the present time $t_0$ is denoted by $a_0$, the light ray reaches the observer at $t_0$. The corresponding proper distance between the observer and the source is then given by $t_{0}$. Thus, $r = a_{0} \int_{t_p}^{t_0} {\frac{dt}{a(t)}}$. This expression measures the particle horizon by accounting for the entire expansion history of the Universe during the propagation of the light signal.

As $t_p$ tends to zero, the particle horizon $R_p$ becomes equal to $r$. Therefore, the particle horizon can be written as $\lim {t_{p} \to 0} \, a_{0} \int_{t_p}^{t_0} {\frac{dt}{a(t)}} = \lim _{z \to \infty} \int_{0}^{z}{\frac{dz}{H(z)}}$. Using Eq. (\ref{e27}), the above expression for the particle horizon can be reformulated via the redshift parameter ($z$) in the following way:
\begin{equation}\label{e38}
R_{p}=\lim_{x \to \infty}\int_0^x \frac{1}{H_0 \left(\frac{\delta +\gamma  (z+1)^{\eta }}{\gamma +\delta }\right)^{\frac{3}{2 \eta }}} \, dz.
\end{equation} 

The particle horizon parameter's nature as the redshift function is plotted and presented in Fig. \ref{fig3}(b) for the different datasets of CC, CMB, Pantheon+SH0ES, Union3.0, and DESI. The analysis of the proposed model indicates that, for the combined dataset {\small(CC+CMB+Union3.0+DESI)}, the particle horizon $R_p$ converges to $3.094^{+0.076}_{-0.066}H_0^{-1}$, as illustrated in Fig. \ref{fig3}(b). Furthermore, Fig. \ref{fig3}(b) reveals that the proper distance $x$ diverges and tends toward infinity as the redshift $z$ approaches zero \cite{Bhardwaj2024lyra,Bentabol2013}. This behavior implies that the event corresponding to the origin of the universe lies at an exceedingly large distance from the present observer, effectively approaching towards infinity.

\subsection{Deceleration parameter (DP)} \label{4.3}
To understand the evolutionary feature of the universe, one can employ the deceleration parameter ($q$), playing a significant role to identify the transition between different cosmic expansion phases. In terms of the Hubble parameter, the deceleration parameter is expressed as $q= -1 + \frac{(1+z)}{H(z)} \frac{d H(z)}{dz} $. Accordingly, the deceleration parameter corresponding to the proposed cosmological model can be written as follows:
\begin{small}
\begin{equation} \label{e39}
q(z) = \frac{3 \gamma  (z+1)^{\eta }}{2 \left(\delta +\gamma  (z+1)^{\eta }\right)}-1.
\end{equation}
\end{small}

However, the combined analysis of the deceleration parameter ($q$) and the Hubble parameter ($H$) provide an extensive highlights of the cosmological expansion dynamics. Key cosmological features, viz. the age of the universe and different expansion phases, can be effectively investigated through these two parameters.

\begin{figure}[ht]
	\centering
	\includegraphics[width=0.9\linewidth]{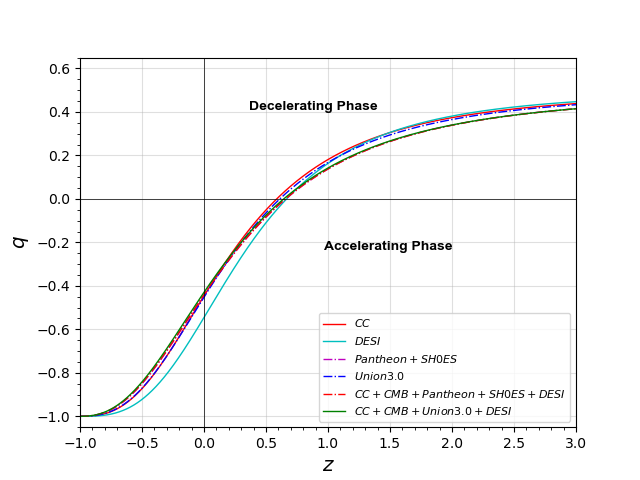}
	\caption{Graphical plot of the deceleration parameter $q$.}\label{fig4}
\end{figure}

The proposed model exhibits a transitional phase of the cosmic expansion history, evolving from an early decelerating phase to the currently accelerating universe, reflecting the dominance of matter in the past and DE at recent epochs. Using the combined CC, CMB, Union3.0, and DESI datasets, the proposed model exhibits a vivid transition within the expansion dynamics and thus yields the present-day value of the DP as $-0.430^{+0.002}_{-0.001}$. The transition from deceleration to acceleration is found to occur at the redshift $z_{t} = 0.643^{+0.052}_{-0.044}$, as illustrated in Fig. \ref{fig4}. Furthermore, the analysis suggests that the accelerated expansion will persist in the future, with $q$ approaching $-1$ as the redshift tends toward $z = -1$.

The numerical values of $q_{0}$ and $z_{t}$ estimated for various datasets are listed in Table \ref{tab2}. The results thus obtained from the proposed model align quite well with the latest experimental findings \cite{Xu2008,Crevecoeur2016,Santos2016,Capozziello2022model,Bhardwaj2025quad}.

\subsection{Statefinders diagnostic} \label{4.4}

To discriminate among different dark energy cosmological models based on their predictive accuracy, two diagnostic quantities known as the statefinder parameters, $r$ and $s$, have been introduced. These parameters provide a more refined characterization of cosmological evolution by tracing the trajectory of a model in the $r—s$ plane. As demonstrated in previous studies \cite{Alam2003,Zhang2005,Setare2007}, the statefinder diagnostic effectively distinguishes various dark energy scenarios. In particular, it offers a clear means of differentiating the proposed cosmological model from the $\Lambda$CDM model through their distinct evolutionary behaviors in the $(r - ~s)$ plane \cite{Alam2003,Sahni2003,Jamil2013}. For the proposed model, the statefinder parameters, viz. $r$ and $s$, can be provided in their standard forms as:
\begin{equation} \label{e40}
r = \frac{2 \delta ^2+2 \gamma ^2 (z+1)^{2 \eta }+\gamma  \delta  (3 \eta -5) (z+1)^{\eta }}{2 \left(\delta +\gamma  (z+1)^{\eta }\right)^2},
\end{equation}
\begin{equation} \label{e41}
s =  -\frac{\gamma  (\eta -3) (z+1)^{\eta }}{3 \left(\delta +\gamma  (z+1)^{\eta }\right)}.
\end{equation}

\begin{figure}[ht]
	\centering
	(a)\includegraphics[width=0.9\linewidth]{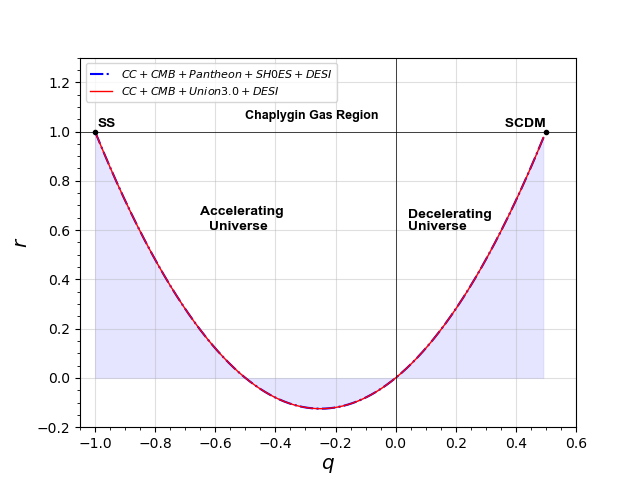}\\
	(b) \includegraphics[width=0.9\linewidth]{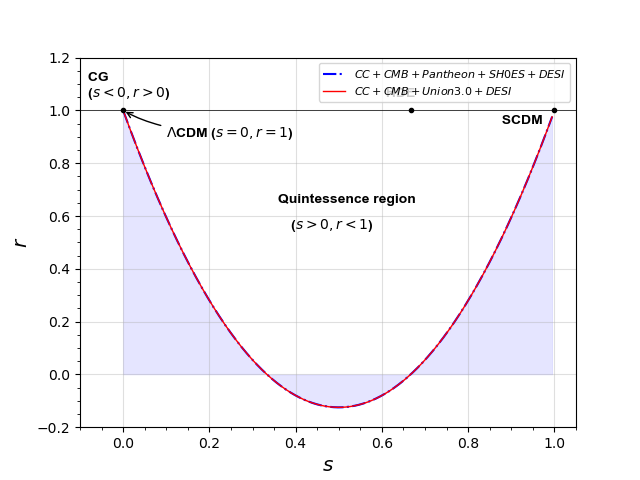}
	\caption{Graphical plots for (a) $r$ vs $q$, (b) $r$ vs $s$.}\label{fig5}
\end{figure}

Figure \ref{fig5}(a) illustrates the evolutionary behavior of the proposed model in the $r—q$ plane. The trajectories clearly indicate a phase transition, as the deceleration parameter $q$ changes from positive to negative values. Initially, the model begins in a matter-dominated epoch corresponding to the SCDM state ($r = 1, \ q = 0.5$). As cosmic evolution proceeds, both $r$ and $q$ decrease until they reach their minimum values. Subsequently, these parameters start increasing and gradually approach the steady-state (SS) point ($r = 1,\ q = -1$). The convergence of the trajectories toward the SS state implies that the proposed dark energy model may exhibit a steady-state-like behavior during the late-time evolution of the Universe.

The distinct dark energy scenarios can be distinguished by their characteristic evolutionary tracks of the geometric pair $(r, ~s)$ in the $r--s$ plane, as illustrated in Fig. \ref{fig5}(b). The trajectories originate away from the SCDM point, ($r = 1, s = 1$), which represents a matter-dominated Cosmos. As cosmic evolution proceeds, these trajectories pass through the quintessence region characterized by ($r < 1, s > 0$) and eventually converge toward the $\Lambda$CDM fixed point, ($r = 1, s = 0$), during the late-time evolution of the Universe.

Therefore, the resulting model exhibits features that closely resembles the $\Lambda$CDM cosmology at the present epoch, although it does not coincide with it exactly.

\subsection{Jerk parameter} \label{4.5}

The jerk parameter ($j$) can be employed to distinguish among different cosmological models. It is motivated by the idea that a sudden change in the expansion dynamics of the universe is required for the transition from a decelerating phase to the presently observed accelerating phase. In cosmological studies, it emerges from the third-order term in the Taylor series expansion of the scale factor $a$ about its present value $a_0$. For the presented model, the jerk parameter can mathematically be defined as \cite{Visser2004}.
\begin{eqnarray} \label{e42}
j&=&1-(1+z) \frac{H'(z)}{H(z)}+\frac{1}{2} \left[(1+z) \frac{ H''(z)}{H(z)}\right]^2,
\end{eqnarray} 
where $ H'(z) = dH(z)/dz$ and $ H''(z) = d^{2}H(z)/dz^2$. For the cosmological model under consideration, the corresponding expression for the jerk parameter (j) can be written as:
\begin{small}
\begin{eqnarray} \label{e43}
&j(z) = \bigg[-\gamma ^4 \left(16 z^2+32 z+25\right) (z+1)^{4 \eta }-4 \gamma ^3 \delta  \left(9 \eta +4 z^2+8 z-5\right) \nonumber\\
&\times (z+1)^{3 \eta }+12 \gamma ^2 \delta ^2 \left(-3 \eta ^2+6 \eta +4 z^2+8 z+1\right) (z+1)^{2 \eta }\nonumber\\
& +80 \gamma  \delta ^3 (z+1)^{\eta +2}+32 \delta ^4 (z+1)^2\bigg] / \bigg[32 (z+1)^2 \left(\delta +\gamma  (z+1)^{\eta }\right)^4\bigg].\nonumber\\
\end{eqnarray}
\end{small}

\begin{figure}[ht]
	\centering
	(a)\includegraphics[width=0.9\linewidth]{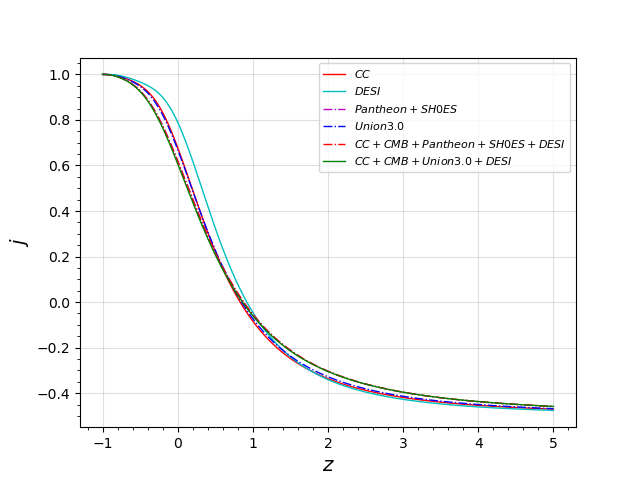}\\
	(b)\includegraphics[width=0.9\linewidth]{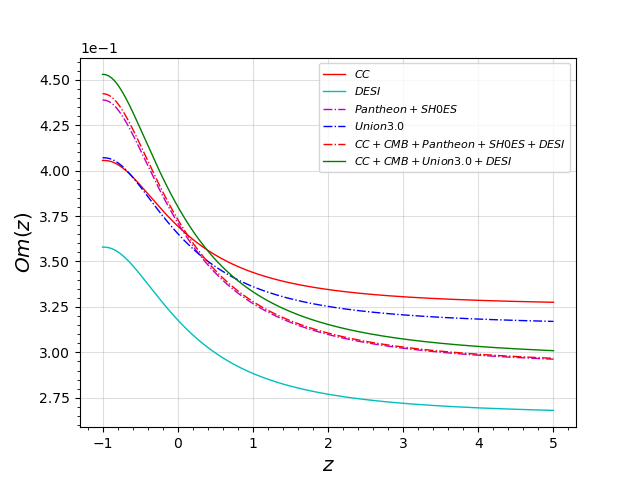}
	\caption{Graphical plots for (a) the behaviour of jerk parameter $ j $ with $ z $, (b) the evaluation of $ Om(z) $. }\label{fig6}
\end{figure}

Figure \ref{fig6}(a) illustrates the evolution of the jerk parameter $j$ as a function of redshift using individual as well as combined observational datasets, including CC, CMB, Pantheon+SH0ES, Union3.0, and DESI. Previous theoretical investigations have indicated that positive values of the jerk parameter are associated with the accelerated expansion of the universe \cite{Blandford2004,Shahalam2015,Bhardwaj2024potenial}. In the present study, the current value of the jerk parameter can be estimated as $0.607^{+0.034}_{-0.028}$ within the joint sample of the CC, CMB, Union3.0, and DESI datasets which clearly supports an accelerating cosmic expansion scenario.

\subsection{$O_{m}$ diagnostic} \label{4.6}

The $O_m$ parameter, first proposed by Sahni et al. \cite{Sahni2008}, is a valuable diagnostic quantity in cosmology. One of its major advantages is that it can be determined directly from the Hubble parameter without requiring derivatives such as $H'(z)$ or additional observational quantities, thereby minimizing the possibility of systematic errors. Owing to its compatibility with both parametric and nonparametric reconstruction techniques, the $O_m$ diagnostic has become widely used in cosmological studies. Moreover, the $O_m$ parameter is capable of distinguishing among different cosmological scenarios even when the matter density parameter and the equation-of-state (EOS) parameter are not known \cite{Jamil2013,Sahni2008}. Following the standard definition presented in Ref. \cite{Sahni2008}, the mathematical expression for $O_m(z)$ for the present model can be given as
\begin{equation} \label{e44}
O_{m}(z)\equiv\frac{\left(\frac{H(z)}{H_0}\right)^2-1}{z^3+3z^2+3 z}  = \frac{-1+\left(\frac{\delta +\gamma  (z+1)^{\eta }}{\gamma +\delta }\right)^{3/\eta }}{z^3+3z^2+3 z}
\end{equation}

The progression of the $O_{m} (z)$ diagnostic is illustrated in Fig. \ref{fig6}(b) for various observational datasets. The behavior of the $O_{m} (z)$ curve plays a significant role in identifying the nature of dark energy models. In particular, a negative slope corresponds to a quintessence-like scenario, whereas a positive slope is associated with phantom-like dark energy. Moreover, a constant $O_{m}(z)$ curve with zero curvature is characteristic of the $\Lambda$CDM cosmological model. The analysis of $O_{m}(z)$ for the present model exhibits a negative slope, indicating quintessence-like behavior and reinforcing the dominance of DE in the current universe. Moreover, the model can be clearly distinguished from the conventional $\Lambda$CDM framework through its positive curvature behavior \cite{Jamil2013,Shahalam2015,Bhardwaj2024f(Q)}.

\section{Energy Conditions} \label{5}

\begin{figure}[ht]
	\centering
	(a)\includegraphics[width=0.9\linewidth]{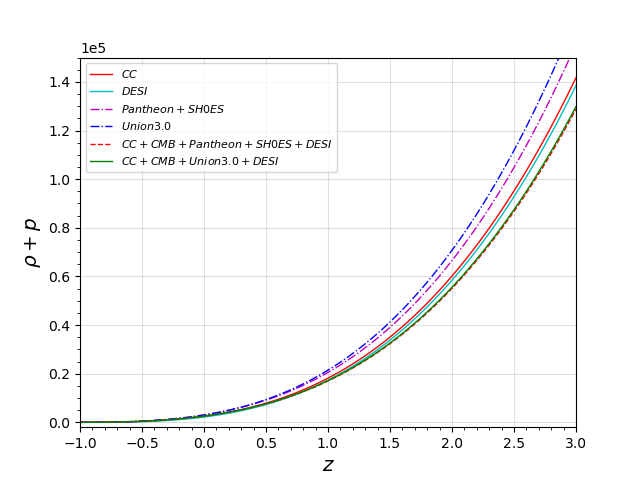}\\
	(b)\includegraphics[width=0.9\linewidth]{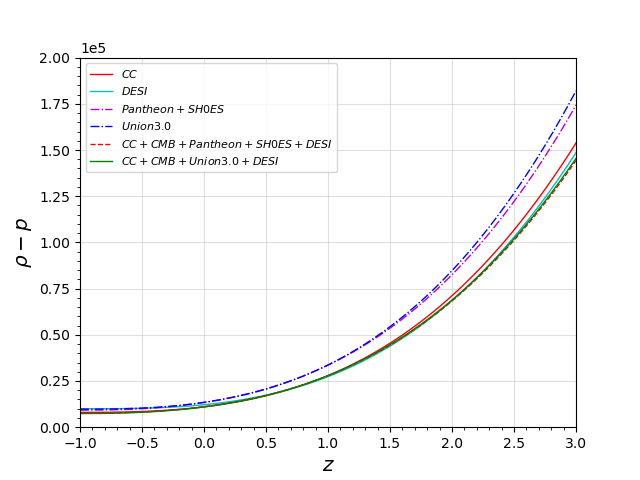}\\
	(c)\includegraphics[width=0.9\linewidth]{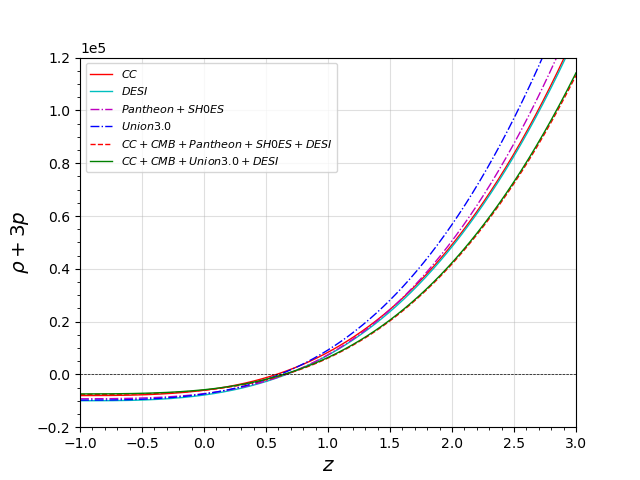}
	\caption{Graphical plots for the behavior of energy conditions.} \label{fig7}
\end{figure}

To assess the physical viability of the proposed cosmological model, we examine the evolution of the energy conditions (ECs) \cite{Carroll2004,Santos2005,Santos2007}. Within the framework of $f(Q,\mathcal{L}_{m})$ gravity, the energy conditions are expressed as follows \cite{Myrzakulov2025energy}: the Weak Energy Condition (WEC) requires $\rho \geq 0$, the Dominant Energy Condition (DEC) is satisfied when $\rho - p \geq 0$, the Null Energy Condition (NEC) demands $\rho + p \geq 0$, and the Strong Energy Condition (SEC) holds for $\rho + 3p \geq 0$.

In general, the Weak Energy Condition (WEC) and Dominant Energy Condition (DEC) are fulfilled by all conventional forms of matter and energy \cite{Bergliaffa2006,Capozziello2018}. On the other hand, the accelerated expansion of the universe is attributed to an exotic energy component, commonly referred to as dark energy (DE), which generates a substantial negative pressure and consequently leads to the violation of the Strong Energy Condition (SEC) \cite{Ratra1988,Bhardwaj2019bulk,Bhardwaj2024f(Q)}.

\begin{figure}[ht]
	\centering
	\includegraphics[width=0.9\linewidth]{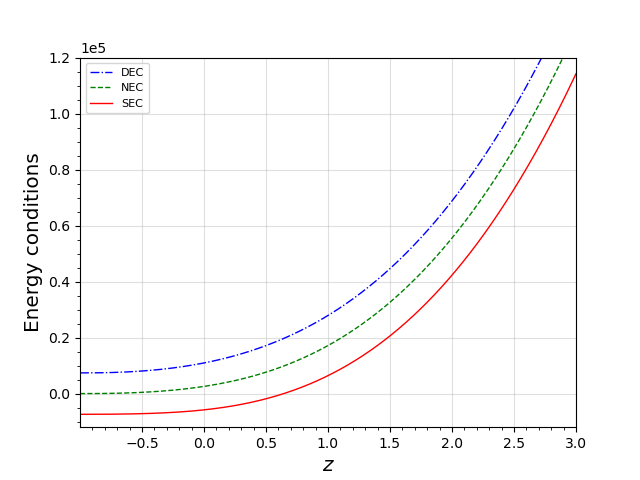}
	\caption{Energy condition for joint dataset of CC+CMB+Union3.0 +DESI.} \label{fig8}	
\end{figure}

Figure \ref{fig7} demonstrates the evolution of energy conditions within the framework of $f(Q,\mathcal{L}_{m})$ gravity theory using various combinations of the CC, CMB, Pantheon+SH0ES, Union3.0, and DESI observational datasets. The analysis reveals that the Dominant Energy Condition (DEC) and Null Energy Condition (NEC) remain satisfied throughout the cosmic evolution, whereas the Strong Energy Condition (SEC) is violated for the proposed model. As shown in Fig. \ref{fig7}, the violation of SEC supports the current accelerated expansion of the Universe. Furthermore, Fig. \ref{fig8} presents the combined behavior of all energy conditions for the proposed model using the joint observational dataset consisting of CC, CMB, Union3.0, and DESI measurements.

\section{Conclusion} \label{6}
In the present work, we have investigated an observationally constrained cosmological model in the framework of $f(Q,\mathcal{L}_{m})$ gravity by considering a suitable parameterization of the Hubble parameter $H(z)$. The modified Friedmann equations were derived for a spatially flat FLRW universe, and the cosmological dynamics of the model were examined in detail. To estimate the free model parameters, we employed recent observational datasets including CC, Pantheon+SH0ES, Union 3.0, DESI BAO, and CMB distance priors using MCMC approach through the $\chi^2$-minimization method. 

The theoretical and numerical analysis demonstrates that the proposed model remains in good agreement with recent cosmological observations. The deceleration parameter exhibits a signature transition from an earlier decelerated phase to the present accelerated epoch, with the transition redshift estimated around $z_t \approx 0.643$. The present value of the deceleration parameter confirms the accelerated nature of the universe. The equation of state parameter remains within the quintessence region and gradually approaches the $\Lambda$CDM limit at late times, indicating consistency with dark-energy-dominated cosmology.

The particle horizon analysis also supports the consistency of the model with the expanding universe scenario. The estimated age of the universe obtained from the joint datasets is approximately $13.724^{+0.087}_{-0.048}$ Gyrs, which is compatible with current observational estimations. Moreover, the Statefinder and Om diagnostic analyses suggest that the proposed model undergoes a transition from a matter-dominated epoch to an accelerated expansion phase characterized by quintessence-like behavior, eventually converging toward the $\Lambda$CDM scenario at late times. The consistently positive values of the jerk parameter throughout the cosmic evolution further support the presence of accelerated expansion. In addition, the examination of the energy conditions shows that the Null Energy Condition (NEC) and Dominant Energy Condition (DEC) remain satisfied, whereas the Strong Energy Condition (SEC) is violated during the cosmic evolution. This violation of the SEC, together with the fulfillment of the NEC and DEC, provides strong evidence for the accelerated expansion of the Universe driven by dark energy.

Overall, the developed $f(Q,\mathcal{L}_{m})$ cosmological model offers an observationally supported and physically viable representation of the late-time accelerating Universe. The obtained results suggest that non-metricity-based modified gravity theories may serve as an effective alternative to explain dark energy and the present cosmic acceleration without invoking exotic matter components.

 \section*{Acknowledgment}
The research of M.K. was carried out in the Southern Federal University with financial support from the Ministry of Science and Higher Education of the Russian Federation (State contract FENW-2026-0028). SR gratefully acknowledges support from the Inter-University Centre for Astronomy and Astrophysics (IUCAA), Pune, India under its Visiting Research Associateship Programme. 

\section*{Declaration of competing interest}
The authors declare that they have no known competing financial interests or personal relationships that could have appeared to influence the work reported in this paper.

 \section*{Data Availability Statement}
No data are generated in this study and we have analyzed publicly available data.

{}
\end{document}